# Silicon as a model ion trap: time domain measurements of donor Rydberg states

N. Q. Vinh[†], P. T. Greenland[‡], K. Litvinenko[§], B. Redlich[†], A. F. G. van der Meer[†],

S. A. Lynch[‡], M. Warner[‡], A. M. Stoneham[‡], G. Aeppli[‡], D. J. Paul[§], C. R. Pidgeon[∥] and

B. N. Murdin[§]*

[†]*FOM Institute for Plasma Physics "Rijnhuizen", P.O. Box 1207, NL-3430 BE*

*Nieuwegein, The Netherlands*

[‡]*London Centre for Nanotechnology and Department of Physics and Astronomy,*

*University College London, WC1H 0AH, England*

[§]*Advanced Technology Institute, University of Surrey, Guildford GU2 7XH, England*

[¶]*Department of Electronics and Electrical Engineering, University of Glasgow, G12 8LT,*

*Scotland*

[∥]*Heriot-Watt University, Department of Physics, Riccarton, Edinburgh, EH14 4AS,*

*Scotland*

*Corresponding author: B N Murdin, Advanced Technology Institute, University of

Surrey, GU2 7XH, England. Fax: +44 (0)1483 689409, E-mail: b.murdin@surrey.ac.uk

(Dated: 29 February 2008)

**ABSTRACT**

One of the great successes of quantum physics is the description of the long-lived Rydberg states of atoms and ions. The Bohr model is equally applicable to donor impurity atoms in semiconductor physics, where the conduction band corresponds to the vacuum, and the loosely bound electron orbiting a singly charged core has a hydrogen-like spectrum according to the usual Bohr-Sommerfeld formula, shifted to the far-infrared due to the small effective mass and high dielectric constant. Manipulation of Rydberg states in free atoms and ions by single and multi-photon processes has been tremendously productive since the development of pulsed visible laser spectroscopy. The analogous manipulations have not been conducted for donor impurities in silicon. Here we use the FELIX pulsed free electron laser to perform time-domain measurements of the Rydberg state dynamics in phosphorus- and arsenic-doped silicon and we have obtained lifetimes consistent with frequency domain linewidths for isotopically purified silicon. This implies that the dominant decoherence mechanism for excited Rydberg states is lifetime broadening, just as for atoms in ion traps. The experiments are important because they represent the first step towards coherent control and manipulation of atomic-like quantum levels in the most common semiconductor and complement magnetic resonance experiments in the literature, which show extraordinarily long spin lattice relaxation times – key to many well-known schemes for quantum computing qubits - for the same impurities. Our results, taken together with the magnetic resonance data and progress in precise placement of single impurities, suggest that doped silicon, the basis for modern microelectronics, is also a model ion trap.

**INTRODUCTION**

Homogenous lifetime broadened two level atoms in ion traps (1) have become favorite objects of study for quantum optics with a view towards both fundamental physics as well as the eventual development of a quantum computer. Among the many schemes proposed (2), the states of ions in trap systems are attractive for the realization of quantum information 'qubits' (quantum bits) because they are well isolated from the decohering effects of the environment and can be coherently controlled by lasers. The Bohr model is equally applicable to donor impurity atoms in semiconductor physics, where the conduction band corresponds to the vacuum, and the loosely bound electron orbiting a singly charged core has a hydrogen-like spectrum according to the usual Bohr-Sommerfeld formula, shifted to the far-infrared due to the small effective mass and high dielectric constant. As with atoms in traps the ground states are tightly confined and well isolated from the environment giving rise to extraordinarily sharp transitions (3),(4),(5) as well as very long spin coherence times (6),(7), measured with magnetic resonance experiments. There are several proposals for quantum information processing based on the spin of silicon donors (8)-(13) and such impurities can now be placed singly with atomic precision (14). In one such scheme (10)-(13), a pair of bismuth impurities are entangled by optical pumping of an adjacent phosphorus atom, strongly coupled by the extended and nearly degenerate excited states. For development of the impurity quantum coherence physics and qubit applications it is crucial to establish time-domain techniques in the relevant frequency range given by the Rydberg in silicon, which is ~ 50 meV rather than 13.6 eV for hydrogen. In particular, the lifetime $T_1$ and decoherence time $T_2$ of the *excited* state of the control impurity must be established, because these set the maximum for the time separation, $t_{sep}$, of the gating pulses (see refs (8) to (11) and (13) for details).

We report, for a Si:P Rydberg state, the first direct measurements of $T_1$ which must, of necessity be performed in the time domain. The required laser power, pulse duration, wavelength coverage, and duty cycle for such measurements are ideally matched to the parameters of the free electron laser FELIX, which gives continuous coverage of the spectral range 5 to 400 meV, controllable pulse durations of between 6 and 100 optical cycles, and peak powers of up to 100MW.

In common with spectroscopy of atoms in gases and traps, frequency domain spectroscopy of the excited states of impurities in semiconductors (see Fig. 1) has a long and distinguished history (17),(18). This remains an active field of research even today, with particular attention given to the extraordinarily narrow linewidths of some of the Rydberg transitions. In the limit of a very clean, homogeneous material, frequency domain spectroscopy provides direct information about the relaxation dynamics. For a real material, however, determination of relaxation times from the frequency domain linewidth is notoriously difficult because the observed shape of the absorption line is generally given by a convolution of the homogenous (or natural) linewidth with the instrument response and a variety of inhomogenous broadening mechanisms. The latter include random strain fields induced by impurities and/or dislocations (18),(19), and other fluctuations in the donor environment due to chemical impurities and different isotopes in the natural composition of Si with differing nuclear moment (3),(4),(5). Time-domain methods such as ours (20),(21) can directly measure the relaxation without any convolution, but require a short pulse laser, in our case a far-infrared free-electron laser. In addition, and much more importantly for future work, they open up the prospect of laser control of impurity states, in precise analogy with the breakthrough that pulsed paramagnetic and nuclear resonance techniques provide relative to continuous wave techniques for electron and nuclear spin resonance. Indeed, the only other time-domain information available for impurities in silicon concerns spin relaxation within the orbital ground state from spin-echo experiments, which have recently been shown to extend to 60 msec for isotopically pure Si:P (6). We therefore chose to study the same donor species.

**TRANSIENT ABSORPTION RESULTS**

Fig. 2 shows the measured probe transmission change as a function of time delay with respect to the pump pulse for the $1s(A_1) \rightarrow 2p_0$ transition at 34.1 meV in P-doped Si. The rise of the leading edge indicates the pulse duration, which was 10 ps in this case. For all pump powers the decay is (almost) exponential; the signal is (almost) linear when plotted on a log-linear scale. Fits with a simple exponential decay gave a value for the lifetime of $T_1 = 205 \pm 18$ ps. This corresponds to a linewidth of $1/T_1 = 0.026$ cm$^{-1}$, i.e. less, but not very much less than the lowest value reported (3) for this transition of 0.034 cm$^{-1}$, which was obtained in an isotopically pure $^{28}$Si sample. Fig. 3 shows the absorption spectrum and lifetimes for P doped Si and As doped Si samples. The absorption spectrum was measured with a resolution of 0.25 cm$^{-1}$ (0.03 meV) at 10 K. The well-known Lyman transitions $1s(A_1) \rightarrow np_0, np_\pm$ between 34 and 45 meV are apparent. No pump-probe effect was seen when the laser was not resonant. The relaxation lifetime of the $2p_0$ state in Si:P has the longest lifetime since it is farthest from the peak in the density of phonon states (22). The fact that the $2p_\pm$ shows a slightly shorter lifetime than the $2p_0$ is also consistent with the spectroscopic linewidth (3). The lifetime of the $2p_0$ state in Si:As is slightly shorter than that of Si:P but we note that in spite of this it is potentially more useful because the energy gap between Rydberg states is larger and overlaps with the available wavelength range of far-infrared semiconductor diode laser pump sources.

**MULTIPHOTON IONISATION**

If we take the absorption cross-section for $1s(A_1) \rightarrow 2p_0$ to be $\sigma_{abs} \sim 3 \times 10^{-18}$ m$^2$ from the small-signal absorption spectrum, and note that the Si transmission coefficient is $\Theta \sim 0.7$ then the lowest photon fluence (the number of photons per unit area integrated over the pulse), $F$, used in Fig. 2 corresponds to a pumping probability of $\Theta F \sigma_{abs} \sim 0.2$, so the excitation densities are small. Photo-ionisation from the excited state is therefore insignificant. However, just as the Rydberg spectrum of the silicon donor impurity is at much smaller energy than for free atoms, so the excited states are much closer to the continuum of conduction band states. The presence of strong multiphoton ionisation processes which have produced a rich variety of atomic and molecular physics effects, but which would interfere with qubit operation, might therefore be expected at higher pump powers relevant for strong-field limit effects such as Rabi oscillations. The photon fluence required for a pulse area $A = \pi$ for $\tau_p = 10$ps is $F \sim 10^{20}$ m$^{-2}$. (The pulse area $A = \mu/\hbar \int E(t)dt$ where $\mu$ is the dipole matrix element and $E(t)$ is the electric field profile of the pulse.) We take the photo-ionisation cross section from the $2p_0$ state to be $\sigma_{ionise} \approx 5 \times 10^{-21}$ m$^2$, calculated by using the hydrogenic $2p \rightarrow$ continuum photo-ionisation cross section, appropriately scaled for the effective mass of Si:P. This results in an ionization probability of $\Theta F \sigma_{\text{ionise}} \sim 0.3$ for the $\tau_p = 10\text{ps}$, $\pi$ pulse so we might expect a small conduction electron population in the strong field limit. We now show that this population is unimportant.

We made measurements similar to those of Fig. 2 up to a maximum fluence of $1.3 \times 10^{20}$ m$^{-2}$, shown in Fig. 4. In all cases we find a decay indistinguishable from a single exponential with the same decay time as in the low power limit. To understand this remarkably simple result, we analyse the dynamics after the pump pulse has passed, i.e. the relaxation and the corresponding recombination of free electrons and ions, with a simple rate equation model (24). Three states are important: i) the ground state, $1s(A_1)$, ii) the state excited by the pump ($2p_0$) and iii) the ionized state, with dimensionless occupation probabilities $n_g$, $n_x$, $n_i$ respectively. Charge conservation implies that the free electron density is equal to the ion density and particle conservation implies $n_g + n_x + n_i = 1$. We have

$$\begin{aligned} \dot{n}_g &= n_x/T_1 + P_g n_i^2 \\ \dot{n}_x &= -n_x/T_1 + P_x n_i^2 \\ \dot{n}_i &= -P_{tot} n_i^2 \end{aligned} \quad [1]$$

The first of these equations represents the feeding of the ground state by decay from the excited state at rate $1/T_1$ and recombination of the electrons and ions with rate $P_g$. This last is proportional to the product of the electron and ion densities, and therefore scales like $n_i^2$. The other equations are similarly interpreted, and $P_{tot} = P_g + P_x$. $P_x$ describes relaxation between the continuum and the excited state which is important at elevated temperature (25). Equations **1** can be integrated analytically [**]$, but inspection

---

[**] Equations **1** can be solved to give

shows that the relaxation gives rise to exponentially decaying terms in the excited population, while the recombination gives rise to reciprocal (*1/t* where *t* is the time after the pump pulse) decays. The recombination rates are given by $P_{g,x} = \sigma_{recom}^{g,x} \bar{v}_e N_0$ where $\sigma_{recom}^{g,x}$ is the cross section for electron capture to the ground or excited state and $\bar{v}_e$ is the mean velocity of the electrons so captured. The recombination time $1/P_{tot} \approx 16$ ps, which sets the time scale for recombination under complete ionisation, can be found by taking Brown's value for the electron recombination cross section $\sigma_{recom} \sim 3 \times 10^{-16} m^2$ (26), and a mean electron velocity $\bar{v}_e \approx 5.3 \times 10^4 ms^{-1}$ appropriate for electrons which have been two-photon ionized. The probe absorption is proportional to $n_g(t) - n_x(t)$. It is clear from equations **1** that $\dot{n}_g - \dot{n}_x$ is unaffected by recombination of free electrons and ions in the case that $P_g \approx P_x$, even if fast. In the case of asymmetric recombination, a fast initial

$$n_g(t) = n_{g0} + n_{x0}\left[1 - e^{-\gamma t}\right] + n_{i0}\left[a(t) - b(t)\right]$$
$$n_x(t) = n_{x0} e^{-\gamma t} + n_{i0} b(t)$$
$$n_i(t) = n_{i0}\left[1 - a(t)\right]$$

where $\gamma = 1/T_1$ and

$$a(t) = \left(\frac{t}{t + t_R}\right)$$
$$b(t) = \left(\frac{P_x}{P_{tot}}\right)\left\{a(t) + e^{-\gamma t} - 1 + \gamma t_R e^{-\gamma(t + t_R)}\left[E_1(-\gamma t_R) - E_1(-\gamma(t + t_R))\right]\right\}$$

In this treatment $E_1(z)$ is the exponential integral $\int_z^\infty \frac{e^{-t}}{t} dt$, and $t_R = \left[n_{i0} P_{tot}\right]^{-1}$ is the initial ion recombination time. The quantities $n_{g0}$, $n_{x0}$ and $n_{i0}$ are the ground state, excited state and ion occupation probabilities produced by the pump pulse.

transient is expected, while the rest of the decay is then dominated by the 200 ps timescale associated with the excited to ground state transition. We see negligible effect of electron-ion recombination on the probe transmission decay, and no initial fast transients, even at the highest pump intensity used (Fig. 4). We conclude that ionization due to multi-photon absorption during pumping is unimportant for interpretation of our experiment, either because we have over-estimated its cross-section or because the recombination is fast and symmetric.

**TEMPERATURE DEPENDENCE**

A key feature distinguishing silicon from ion traps is that there is a bath of excitations – Si lattice phonons – coupled to the donor levels. Warming will increase the population of the bath, eventually causing depopulation and decoherence. To quantify this effect and determine the temperature range over which Si can meaningfully function as an ion trap, we have measured the temperature dependence of the $2p_0$ and $2p_\pm \to 1s(A_1)$ decay times. Fig. 5 shows the remarkable result that the decay times actually increase with temperature - more obviously for $2p_\pm$ than for $2p_0$ - displaying a maximum at 50K. We explain this temperature dependence with a phenomenological equation for the effective relaxation time (27):

$$\frac{1}{\tau_{eff}(T)} = \frac{1}{T_1} - R_a e^{-\Delta E_a/kT} + R_b e^{-\Delta E_b/kT} \qquad [2]$$

The first term on the right describes the direct population relaxation from $2p_0$ to $1s(A_1)$. The second term, given earlier for ionized acceptors by Cuthbert (27), comes about because raising the temperature increases the number of equilibrium free electrons. The upshot is an increased effective lifetime as measured in our absorption experiment, which senses a recovery of the $1s \rightarrow 2p_0$ signal only when the $1s$ state is reoccupied, which is, of course, less likely when the original electrons are far from the donors. At higher temperatures the Boltzmann tail of the free electron distribution can have enough energy to enable thermal excitation sufficiently far into the conduction band for subsequent recombination via emission of the strong transverse optical phonon, energy $E_{TO} \approx 60 meV$. This gives rise to the third term. At the same time, of course, the Saha equation (28) predicts that the equilibrium density of the ground state disappears on a similar energy scale which is much smaller than that (~ 10,000K) for hydrogen.

The insert of Fig. 5 illustrates the effects described above, and shows how the adjustable parameters of equation **2** can be interpreted: $\Delta E_a$ is the ionization energy for $2p$ electrons, $\Delta E_b = E_{TO} - E_{21}$ is the energy required to activate the $2p$ electrons to a state whence they may decay to the $1s(A_1)$ ground state by optical phonon emission; $1/R_b = 1/R_{TO}$ is the optical phonon emission lifetime and $E_{21}$ is the energy of $2p \rightarrow 1s(A_1)$ transitions. The solid line in Fig. 5 corresponds to the following values of the fitting parameters: $T_1 = 215 \pm 10 ps$, $\Delta E_a = 11.8 \pm 1.1 meV$, $\Delta E_b = 32.1 \pm 2.1 meV$ meV and $1/R_b = 1.7 ps$. The energy values for the excited state and the ground state involved in the relaxation process for the $2p_0 \rightarrow 1s(A_1)$ transition are in good agreement with $E_a = 11.5 meV$, $E_{TO} - E_{21} \approx 30 meV$. The lifetime of 215 ps is also close to the 205 ps found above. The temperature dependence of the lifetime of the $2p_\pm$ is similar to that of the $2p_0$ because the transition energy is similar, though the ionization energy is smaller by a factor of two giving rise to the steeper initial increase. Numerical solution of the rate equations to finite temperature, including statistical detailed balance, confirms our interpretation. We remark that, even though recombination is an important process at higher temperature, the probe transmission still shows an approximately exponential time dependence. It is also worth noting that the agreement of the simple model above with the data again indicates that multiphoton ionization by the pump is unimportant for our results.

**CONCLUSIONS**

In summary, we have shown that, impurities in Si share significant virtues with isolated atoms in traps, though phonon, rather than photon emission leads to decay timescales 10 to 100 times faster than what is usual in atoms. Our time-domain data show directly that population decay effects are the dominant contribution to frequency domain linewidths of Rydberg levels in isotopically pure silicon. Comparing with linewidths from the frequency domain, we find that, at low temperature in isotopically purified material, the dominant decoherence mechanism is *lifetime* broadening due to the emission of phonons. The donors can be effectively isolated from the environment and have no significant sources of decoherence other than population decay by emission of phonons. In addition, we show remarkable insensitivity of the results to multi-photon effects as we vary the power of the intense free electron laser beam, and discover that initially, the recovery time for $2p \to 1s$ absorption actually increases with temperature. Finally, $T_1$ is 6 times the Larmor precession period for the $g = 2$ impurity spins in Si at the modest external field of 1 tesla, implying the possibility of programming sequences of spin and Rydberg state operations on impurities in the world's best understood material, as required for proposed quantum qubit schemes (8)-(13).

**EXPERIMENTAL PROCEDURE**

We performed a pump-probe measurement of the lifetimes for different photon energies resonant with transitions between the silicon impurity Rydberg states using the FELIX free-electron laser at the FOM Institute for Plasma Physics "Rijnhuizen", Nieuwegein, The Netherlands. In this technique, described in detail elsewhere, a strong pump pulse causes bleaching of the absorption measured by the weak probe pulse, the recovery of which is then measured as a function of the delay between the two (20),(21). The samples investigated were float-zone grown Si wafers of thickness ~200 μm and doped with P or As to donor concentrations of $N_0 \sim 2 \times 10^{21} \text{m}^{-3}$. The silicon was 'natural', *i.e. not* isotopically purified.

**ACKNOWLEDGEMENTS**

We gratefully acknowledge the support of EPSRC and the Stichting voor Fundamenteel Onderzoek der Materie (FOM) in providing the required beam time on FELIX. This work was funded by the EC IST FET programme SHINE (IST-2001-38035), EPSRC grant GR/S23506, EPSRC Advanced Fellowship EP/E061265/1, the Basic Technologies Programme of RCUK, and the Royal Society Wolfson Research Merit Award Scheme.

**FIGURE CAPTIONS**

Fig. 1. Rydberg series and optical transitions from the lowest energy states of the hydrogen-like donor impurities phosphorus and arsenic in silicon (17),(18).

Fig. 2. The change in probe transmission induced by the pump as a function of the time delay between pump and probe, observed in the Si:P sample for the $1s(A_1) \rightarrow 2p_0$ transition at a sample temperature, T, of 10 K and a pump and probe photon energy of 34.1 meV. The rise of the leading edge indicates the pulse duration, which was ~10 ps. The laser pump powers used corresponded to micropulse energies shown on the figure. The lowest pump pulse energy (1.1nJ) corresponds to a focussed photon fluence of ~ $10^{17}$ photons m$^{-2}$. Also shown are fits using a single exponential decay where the decay parameter is the spontaneous relaxation rate $1/T_1$. The inset shows the transient pump-probe experimental geometry.

Fig. 3. (a) The absorption spectrum for P-doped Si measured by FTIR spectroscopy with 0.25 cm$^{-1}$ or 0.03 meV resolution. The sample temperature was 10 K. The lifetimes of the indicated states, determined from pump-probe signals (such as those in Fig. 2) are also shown. (b) The corresponding results for Si:As. (c) The one-phonon density of states of silicon, which, of course, determines the phonon emission decay rate at low temperature (taken from Ref. (23)).

Fig. 4. The probe transmission at high pump intensity is shown as a function of pump-probe delay, along with a fitted single exponential decay and a reciprocal decay. We plot both the logarithm (upper panel) and reciprocal (lower panel) of the signal. If spontaneous decay is most important we expect the former to be linear as a function of time; if recombination from the conduction band is most important, then the latter will show linear behaviour. At high intensities, even though two-photon ionization is likely to be strong, the experimental signal is exponential.

Fig. 5. The temperature dependence of the lifetimes for $1s(A_1) \to 2p_0, 2p_\pm$ transitions at 34.1 and 39.2 meV, respectively in P-doped Si. The lines are fits of equation **2**. The inset shows the level scheme with transitions.

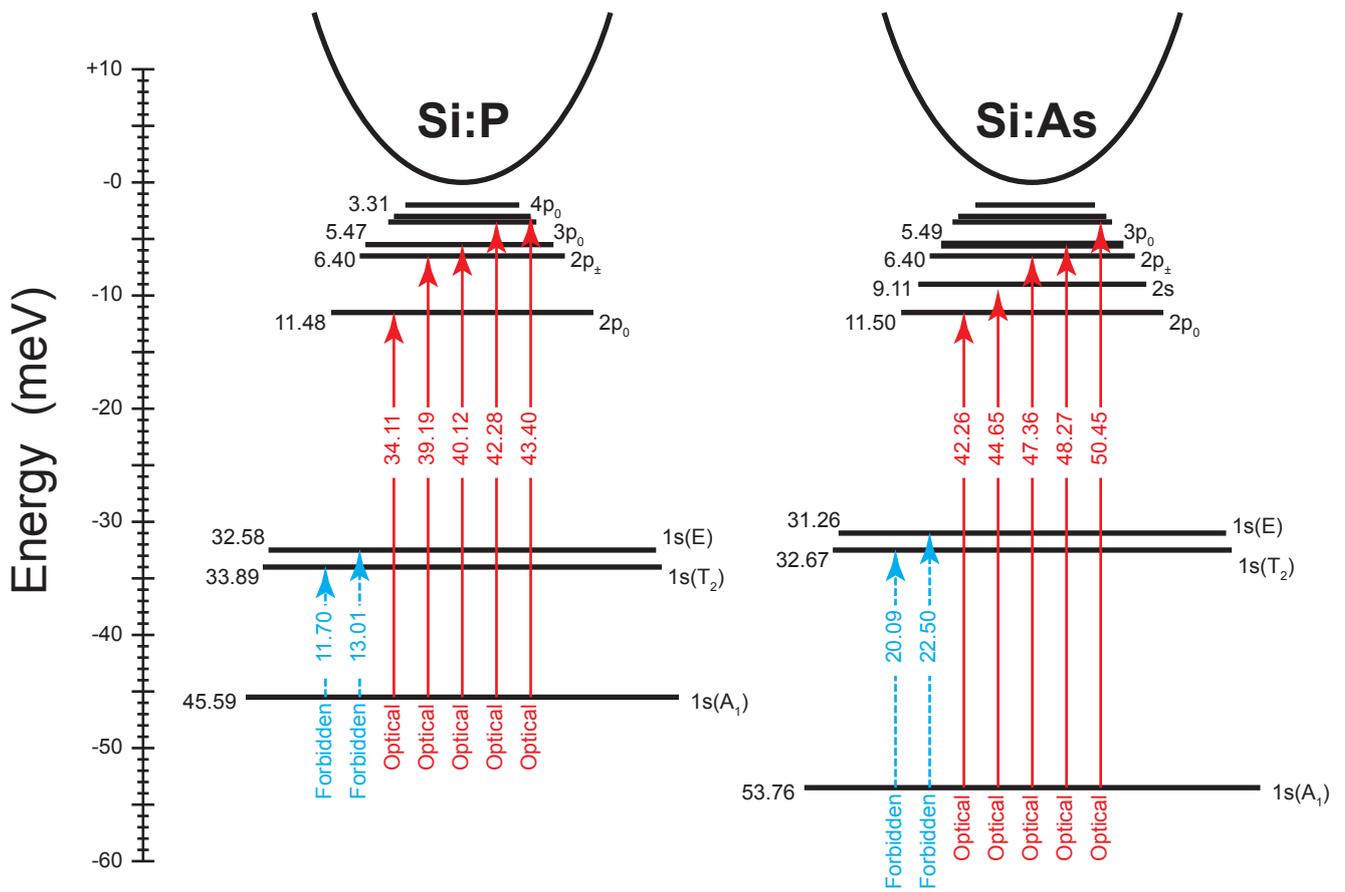

Figure 1

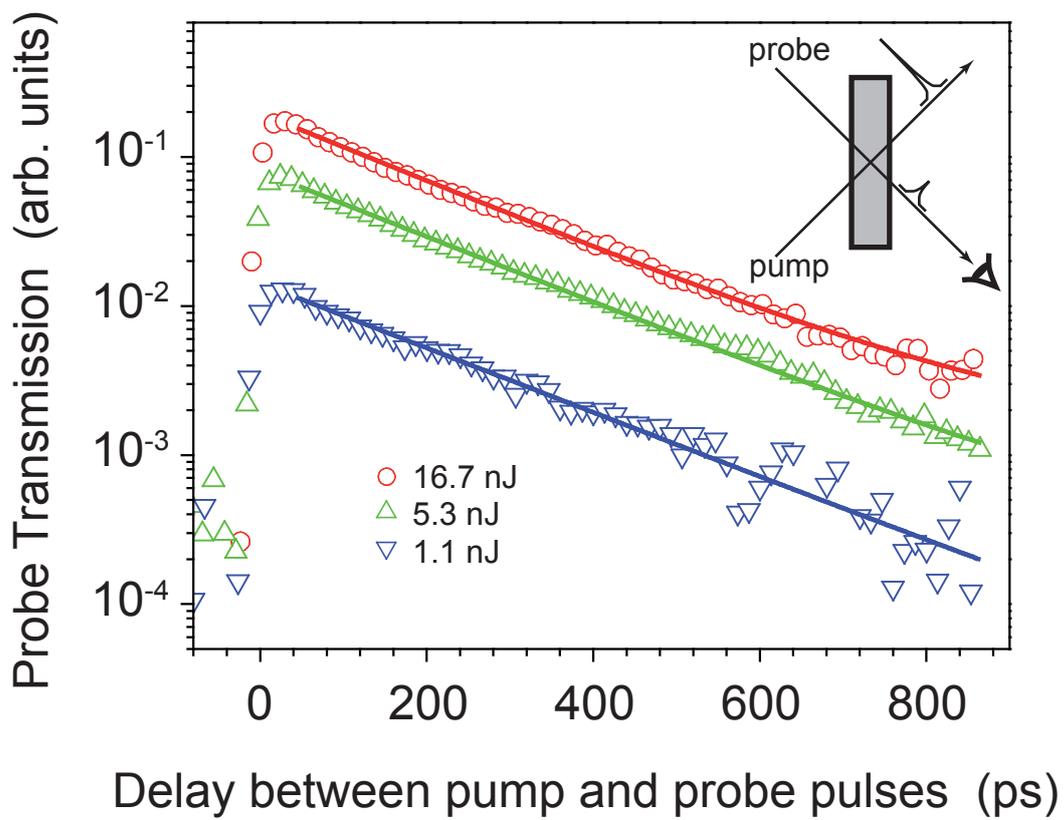

Figure 2

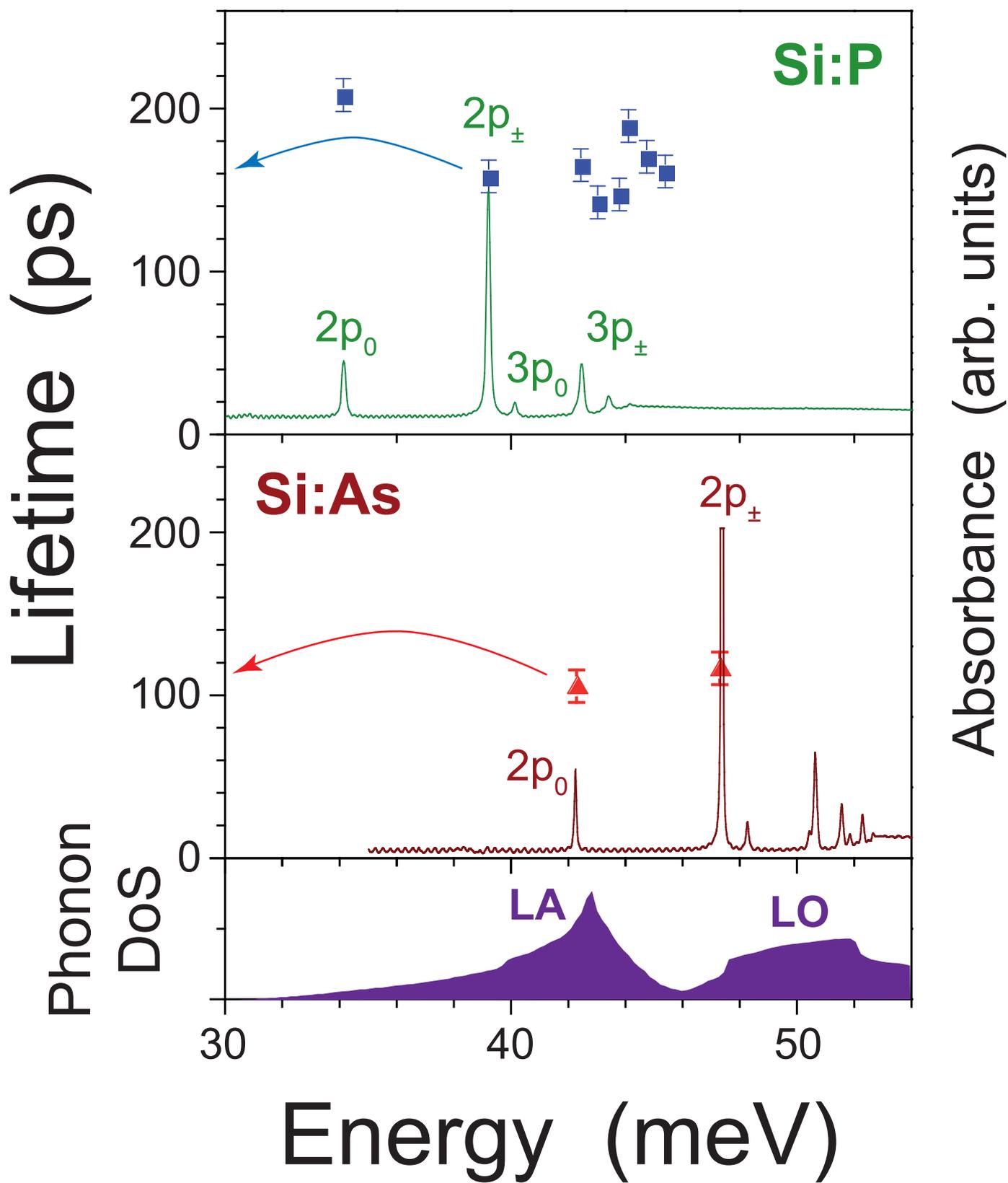

Figure 3

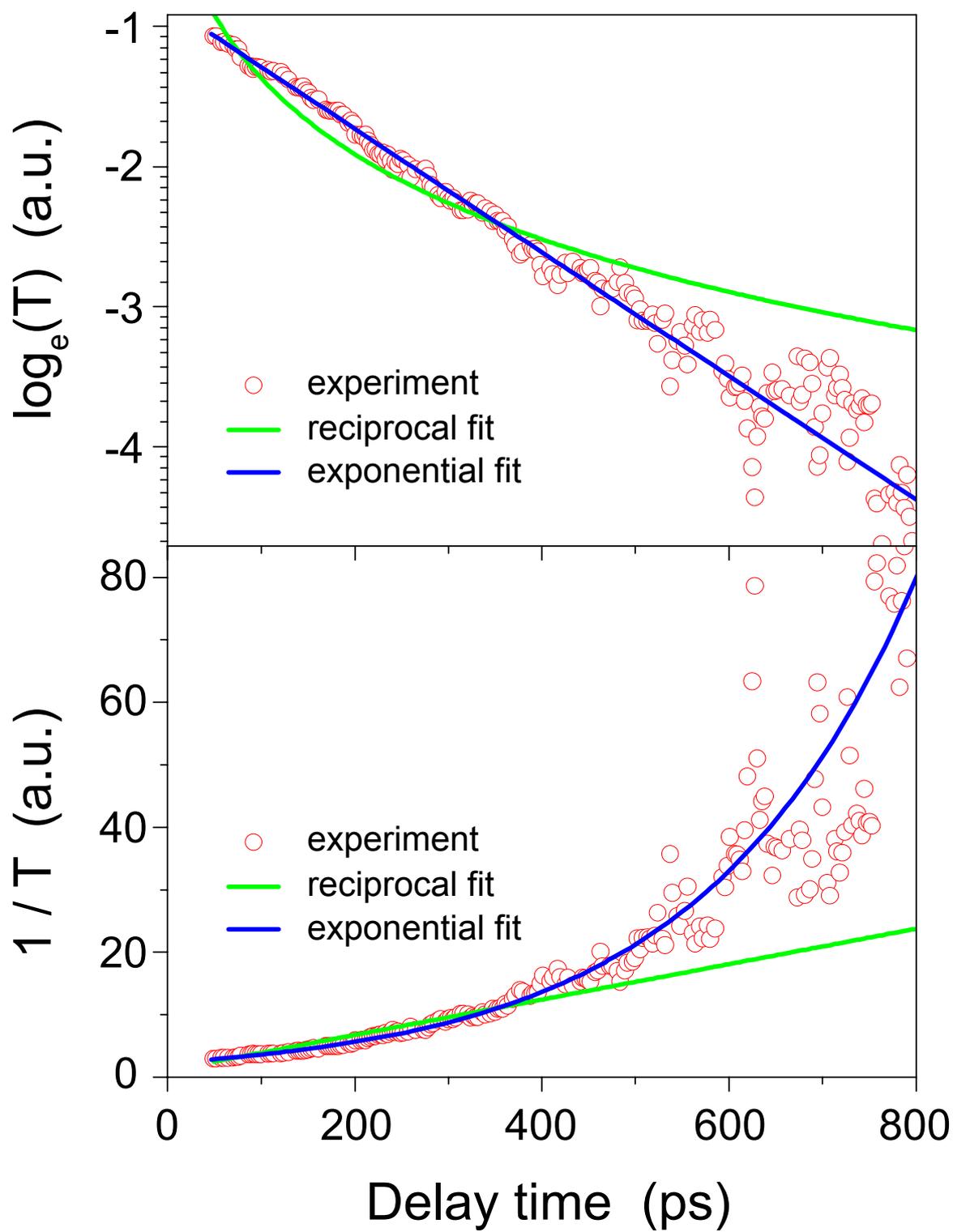

Figure 4

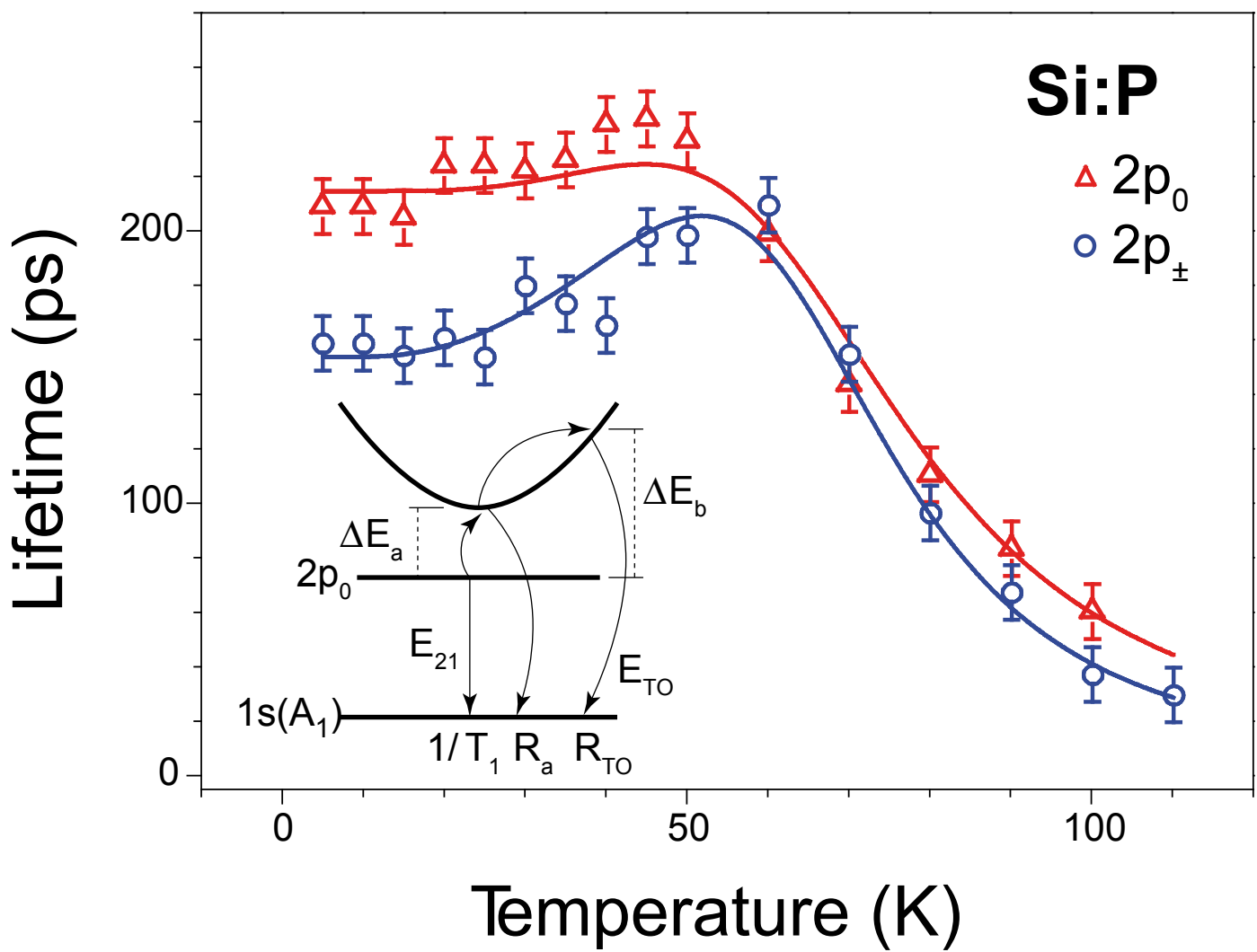

Figure 5